\documentclass[12pt,prd,tightenlines,nofootinbib,setspace]{revtex4}
\usepackage{bm}
\usepackage{graphics}
\usepackage{epsfig}
\usepackage{amssymb}
\usepackage{amsmath}
\usepackage{amsthm}
\usepackage{amsfonts}
\usepackage{amssymb}
\usepackage{braket}
\usepackage{rotating}

\begin{document}

\title{Low-lying electron energy levels in three-particle\\ 
electron-muon ions $(\mu e Li)$, $(\mu e Be)$, $(\mu e B)$}

\author{A.~E.~Dorokhov,\footnote{E-mail:~dorokhov@theory.jinr.ru}}
\author{V.~I.~Korobov,\footnote{E-mail:~korobov@theory.jinr.ru}}
\affiliation{Joint Institute of Nuclear Research, BLTP, \\
141980, Moscow region, Dubna, Russia}
\author{A.~P.~Martynenko,\footnote{E-mail:~a.p.martynenko@samsu.ru}}
\author{F.~A.~Martynenko\footnote{E-mail:~f.a.martynenko@gmail.com}}
\affiliation{Samara National Research University named after acad. S.P. Korolev, \\
443086, Samara, Russia}

\begin{abstract}
The electron Lamb shift $ (2P-2S) $ and the energy interval $ (2S-1S) $ in 
muon-electron ions of lithium, beryllium, and boron have been calculated within 
the framework of the perturbation theory method for the fine structure constant 
and the electron-muon mass ratio. The corrections of the first and second orders 
of the perturbation theory, which include the effects of vacuum polarization, 
nuclear structure and recoil, are taken into account. The obtained analytical results in 
perturbation theory are compared with the results of calculations within 
the variational approach. The obtained values for the Lamb shift 
$ (2P-2S)$ and the interval $(2S-1S) $ can be used for comparison with future 
experimental data and verification of quantum electrodynamics.
\end{abstract}

\maketitle

\section{Introduction}

Muon-electron ions of lithium, beryllium and boron $ (\mu e Li) $, $ (\mu e Be) $, 
$ (\mu e B) $ are the simplest three-particle systems, consisting of a nucleus, 
a negatively charged muon and electron. The main interaction in this system is 
determined by the Coulomb interaction of charged particles. Bound particles in 
muonic lithium, beryllium and boron ions have different masses 
$ m_e \ll m_\mu \ll M $. As a result, the muon and the nucleus form a pseudonucleus, 
and in the first approximation, the muon-electron ions of lithium, beryllium and boron 
can be considered as two-particle systems. Such three-particle systems are interesting 
in that they allow one to study in the energy spectrum simultaneously the corrections 
of small distances associated with the motion of the muon as in muonic two-particle 
atoms (to the structure of the nucleus and the polarization of the vacuum) and the corrections 
of large distances associated with the motion of an electron as in electron atoms.

There are two ways to calculate the energy levels in three-particle muon-electron 
atoms and ions. The first approach used in \cite{lm1,lm2,ya,ya1,km1,km2,km3,sgk},
is based on the perturbation theory (PT) method for the Schr\"odinger equation. 
In this case, there is an analytical solution for the three-particle wave function 
in the initial approximation, which allows one to take into account various 
corrections to energy levels from other interactions according to the perturbation 
theory. In another approach, \cite{hh,Chen,hh1,d,frolov,korobov,korobov1} used 
a variational method in quantum mechanics to find the energy levels of three particles. 
It made it possible to obtain numerical values of the energy levels of a three-particle system with very 
high accuracy. 

Precision muonic physics has become especially important
since 2010, when the first experimental results on the measurement of low-lying energy levels of muonic 
hydrogen were obtained by the CREMA (Charge Radius Experiments with Muonic Atoms) collaboration \cite{pohl}. 
A decade of active work of this collaboration brought interesting and unexpected results, related primarily 
to the determination of more accurate values of the charge radii of light nuclei (proton, deuteron, helion, alpha particle).
The CREMA experiments have caused a whole series of new experimental studies of the
muonic systems.
Physics of muonic two-particle and three-particle systems remains an urgent problem that requires 
appropriate theoretical studies and calculations of the observed quantities with high accuracy.

The purpose of this work is to calculate the electron 
Lamb shift (2S-2P) and the energy interval $ (2S-1S) $ as in the framework of the first approach from 
\cite{km1,km2,km3} for electron-muonic ions with a nuclear charge of 3, 4, 5 
and within the framework of the variational method. Note that the hyperfine structure 
of muonic helium $ (\mu e He) $ was measured in \cite{gardner}. 
New plans for precision microwave spectroscopy of the J-PARC MUSE collaboration \cite{strasser}
are related to the measurement 
of the hyperfine structure of the ground state of muonic helium with an accuracy two orders of magnitude 
higher than the accuracy of previous experiments 1980s.
Measurement of other energy intervals, such as $ (2S-1S) $, $(2S-2P) $, in muon-electron helium 
or muon-electron ions of lithium, beryllium and boron is quite feasible.

\section{General formalism}

The Hamiltonian of the three-particle system muon-electron-nucleus has the 
following general structure \cite{km1,km2,km3}:
\begin{equation}
\label{eq1}
H=H_0+\Delta H+\Delta H_{rec}+\Delta H_{vp}+\Delta H_{str},
\end{equation}
\begin{equation}
\label{eq2}
H_0=-\frac{1}{2M_\mu}\nabla^2_\mu-\frac{1}{2M_e}
\nabla^2_e-\frac{Z\alpha}{x_\mu}-\frac{(Z-1)\alpha}{x_e},
\Delta H=\frac{\alpha}{x_{\mu e}}-\frac{\alpha}{x_e},~\Delta H_{rec}=-\frac{1}{M}
{\mathstrut\bm\nabla}_\mu\cdot{\mathstrut\bm\nabla}_e,
\end{equation}
where $Z$ is the nucleus charge, ${\bf x}_\mu$, ${\bf x}_e$ are the muon and electron
radii vectors with respect to the nucleus, $x_{\mu e}=|{\bf x}_\mu-{\bf x}_e|$,
$M_e=m_eM/(m_e+M)$ and $M_\mu=m_\mu M/(m_\mu+M)$ are the reduced masses
of electron-nucleus and muon-nucleus subsystems. The Hamiltonian terms $\Delta H_{vp}$,
$\Delta H_{str}$ and $\Delta H_{rec}$ determine the vacuum polarization, nuclear
structure and recoil corrections. In the initial approximation the wave function
of the three-particle system with muon in the ground state and electron in the
$1S$, $2S$ or $2P$ states takes the form:
\begin{equation}
\label{eq3a}
\Psi_{1S}({\bf x}_\mu,{\bf x}_e)=
\frac{1}{\pi}(W_\mu W_e)^{3/2}e^{-W_\mu x_\mu}e^{-W_e x_e},
\end{equation}
\begin{equation}
\label{eq3}
\Psi_{2S}({\bf x}_\mu,{\bf x}_e)=
\frac{1}{2\sqrt{2}\pi}(W_\mu W_e)^{3/2}(1-\frac{1}{2}W_e x_e)e^{-W_\mu x_\mu}
e^{-\frac{1}{2}W_e x_e},
\end{equation}
\begin{equation}
\label{eq4}
\Psi_{2P}({\bf x}_\mu,{\bf x}_e)=
\frac{1}{2\sqrt{6\pi}}(W_\mu W_e)^{3/2}W_ex_e({\mathstrut\bm\varepsilon}{\bf n})
e^{-W_\mu x_\mu}e^{-\frac{1}{2}W_e x_e},
\end{equation}
where $W_e=(Z-1)M_e\alpha$, $W_\mu=ZM_\mu\alpha$.
The wave function of the $2P$ state is presented in tensor form
and ${\mathstrut\bm\varepsilon}$ is the polarization vector of the state
$2P$.
The main contribution to the perturbation operator is determined by the term 
$ \Delta H $. It is known that in an electron hydrogen-like atom the Lamb shift 
$ 2S-2P $ is a radiation effect of the fifth order in the fine structure constant 
$ \alpha $, and in a muonic hydrogen-like atom the Lamb shift is determined 
by the effect of vacuum polarization in the leading order $\alpha^3 $. 
The purely Coulomb interaction of charged particles does not give a shift 
between the $ 2S$ and $ 2P$ levels in two-particle atoms. In three-particle 
systems, the electron Lamb shift in the leading order $\alpha^2$ is determined 
by the purely Coulomb interaction.

In the initial approximation, the energy of the system is equal to the sum of 
the Coulomb energies of an electron and a muon of the order of $O(\alpha^2) $. 
So, for example, if both the electron and the muon are in the $ 1S $ state, then this 
energy is equal to $[-\frac{1}{2}M_e(Z-1)^2 \alpha^2-\frac{1}{2}M_\mu Z^2 \alpha^2]$. 
In what follows, the muon energy is not of interest to us, 
since it will cancel out in the $ (2P-2S)$ and $ (2S-1S) $ intervals. In the 
case of the Lamb shift $(2P-2S)$ in the first order of the perturbation theory, 
it is necessary to calculate two matrix elements of the Coulomb interaction 
$\Delta H $:
\begin{equation}
\label{eq5}
\begin{array}{@{}l}\displaystyle
\Delta E^{(1)}(2S) =
   \left\langle
      \Psi_{2S}\left|\left(\frac{\alpha}{x_{\mu e}}-\frac{\alpha}{x_e}\right)\right|\Psi_{2S}
   \right\rangle,
\\[4mm]\displaystyle
\Delta E^{(1)}(2P) =
   \left\langle
      \Psi_{2P}\left|\left(\frac{\alpha}{x_{\mu e}}-\frac{\alpha}{x_e}\right)\right|\Psi_{2P}
   \right\rangle.
\end{array}
\end{equation}

For the $2S$ electron state the matrix element has the form ($a_1=W_e/W_\mu$):
\begin{equation}
\label{eq6}
\begin{array}{@{}l}\displaystyle
\Delta E^{(1)}(2S) =
   \frac{(W_eW_\mu)^3}{8\pi^2}\int d{\bf x}_e d{\bf  x}_\mu
   \left(1-\frac{1}{2}W_ex_e\right)^2e^{-2W_\mu x_\mu} e^{-W_e x_e}
\left(\frac{\alpha}{|{\bf x}_\mu-{\bf x}_e|}-\frac{\alpha}{x_e}\right)
\\[4mm]\displaystyle\hspace{10mm}
=\alpha W_e\left[-\frac{1}{4}+\frac{8\!+\!a_1(20\!+\!a_1(12\!+\!a_1(10\!+\!a_1)))}{(2+a_1)^5}\right]=
W_e\alpha\left(-\frac{1}{4}a_1^2+\frac{5}{8}a_1^3-\frac{63}{64}a_1^4+\dots\right).
\end{array}
\end{equation}

Averaging over the orbital angular momentum projections by means of the relation
\begin{equation}
\label{lll}
\frac{1}{3}\sum_\lambda\varepsilon_i^{\ast(\lambda)}\varepsilon_j^{(\lambda)}=
\frac{1}{4\pi}\delta_{ij},
\end{equation}
the matrix element for the $2P$ electron state can be calculated analytically in the same way:
\begin{equation}
\label{eq7}
\begin{array}{@{}l}\displaystyle
\Delta E^{(1)}(2P)=
\frac{(W_eW_\mu)^3}{96\pi^2}\int d{\bf x}_e d{\bf  x}_\mu
(W_ex_e)^2e^{-2W_\mu x_\mu} e^{-W_e x_e}
\left(\frac{\alpha}{|{\bf x}_\mu-{\bf x}_e|}-\frac{\alpha}{x_e}\right)
\\[4mm]\displaystyle\hspace{10mm}
=\alpha W_e\left[-\frac{1}{4}+\frac{((2+a_1)^5-6a_1^4-a_1^5)}{4(2+a_1)^5}\right]=
W_e\alpha\left(-\frac{3}{64}a_1^4+\frac{7}{64}a_1^5+\dots\right),
\end{array}
\end{equation}

and an expansion in $a_1$ converges well, because numerical values of $a_1$ for muonic
ions of lithium, beryllium and boron are small:
$a_1(\mbox{Li})=0.003276$, $a_1(\mbox{Be})=0.003673$, $a_1(\mbox{B})=0.003909$.
The parameter $a_1=(Z-1)M_e/ZM_\mu$ determines the recoil effects
and acts as a small parameter used in the framework of the perturbation theory
for the $ \Delta H$ interaction.
We see that the recoil corrections have a higher degree of smallness for the $2P$
state than for the $ 2S$. The main contribution to the electron Lamb shift is of
order $ O(\alpha^2)$:
\begin{equation}
\label{eq8}
\Delta E^{(1)}_{LO}(2P-2S)\approx W_e\alpha\>\frac{a_1^2}{4}=\frac{M_e\alpha^2}{4}
\frac{(Z-1)^3M_e^2}{Z^2M_\mu^2}.
\end{equation}

Thus, the electron Lamb shift occurs in such three-particle systems in the order
$O(\alpha^2)$ due to the Coulomb interaction of all particles. Note that the
recoil correction $ \langle\Psi|\Delta H_{rec}|\Psi\rangle$ is equal to 0 for both
states of the electron.
The refinement of the \eqref{eq8} result is related to the inclusion of corrections
in the higher orders of the perturbation theory in $ \alpha $ and $ m_e /m_\mu $.
In the second order of perturbation theory, the correction to energy levels is
determined by the following expression:
\begin{equation}
\label{eq9}
\begin{array}{@{}l}\displaystyle
\Delta E^{(2)}=\int \psi_{\mu 0}({\bf x}_\mu)\psi_{e1}({\bf x}_e)
\left(\frac{\alpha}{x_{\mu e}}-\frac{\alpha}{x_e}\right)
\sum_{n,n'}\frac{\psi_{\mu n}({\bf x}_\mu)\psi_{en'}({\bf x}_e)\psi_{\mu n}({\bf x'}_\mu)
\psi_{en'}({\bf x'}_e)}{E_{\mu 0}+E_{e1}-E_{\mu n}-E_{en'}}
\\[3mm]\displaystyle\hspace{20mm}
\times\>\psi_{\mu 0}({\bf x'}_\mu)\psi_{e 1}({\bf x'}_e)
\left(\frac{\alpha}{x'_{\mu e}}-\frac{\alpha}{x'_e}\right)d{\bf x}_\mu
d{\bf x'}_\mu d{\bf x}_e d{\bf x'}_e,
\end{array}
\end{equation}
where $\psi_{\mu 0}({\bf x}_\mu)$ is the muon wave function in the ground state,
$\psi_{e1}({\bf x}_e)$ is the electron wave function in states $2S$, $2P$
(or $1S$).
The reduced Coulomb Green's function entering \eqref{eq9} is determined by the sum 
over the excited muonic and electronic states. Let's split the complex matrix element 
\eqref{eq9} into several simpler ones, highlighting certain states of the muon. 
Let the muon be in an intermediate state $ n = 0 $. Then from \eqref{eq9} we get 
the first part of the correction:
\begin{equation}
\label{eq10}
\begin{array}{@{}l}\displaystyle
\Delta E_1^{(2)}=\int \psi_{\mu 0}({\bf x}_\mu)\psi_{e1}({\bf x}_e)
\left(\frac{\alpha}{x_{\mu e}}-\frac{\alpha}{x_e}\right)
\psi_{\mu 0}({\bf x}_\mu)\psi_{\mu 0}({\bf x'}_\mu)
\sum_{n'}\frac{\psi_{en'}({\bf x}_e)\psi_{en'}({\bf x'}_e)}{E_{e1}-E_{en'}}
\\[3mm]\displaystyle\hspace{20mm}
\times\>\psi_{\mu 0}({\bf x'}_\mu)\psi_{e 1}({\bf x'}_e)
\left(\frac{\alpha}{x'_{\mu e}}-\frac{\alpha}{x'_e}\right)
d{\bf x}_\mu d{\bf x'}_\mu d{\bf x}_e
d{\bf x'}_e.
\end{array}
\end{equation}

The reduced Coulomb Green's function of an electron for an excited state with 
$ n=2 $ has two parts. Only the part $ \tilde G^e_ {2S}$ gives a nonzero 
contribution to this matrix element, which has the following form \cite{hameka}:
\begin{equation}
\label{eq11r}
\begin{array}{@{}l}\displaystyle
\widetilde G^e_{2S}(r_1,r_2)=-\frac{(Z-1)\alpha M_e^2}{16\pi x_1x_2}e^{-\frac{1}{2}(x_1+x_2)}
g_{2S}(x_1,x_2),
\\[3.5mm]
g_{2S}(x_1,x_2)=8x_<-4x^2_<+8x_>+12x_<x_>-26x^2_<x_>+2x^3_<x_>-4x^2_>-26x_<x^2_>
\\[3mm]\hspace{20mm}
+23x^2_<x^2_>-x^3_<x^2_>
%\\[3mm]\hspace{20mm}
+2x_<x^3_>-x^2_<x^3_>+4e^{x_<}(1-x_<)(x_>-2)x_>
\\[3mm]\hspace{20mm}
+4(x_<-2)x_<(x_>-2)x_>\bigl[-2\gamma+\mathrm{Ei}(x_<)-\ln(x_<)-\ln(x_>)\bigr],
\end{array}
\end{equation}
where $x_<=min(x_1,x_2)$,  $x_>=max(x_1,x_2)$, $x_i=Wr_i$, $\gamma$ is the Euler 
constant. Simple matrix elements over the muon coordinates are calculated analytically:
\begin{equation}
\label{eq12}
V_\mu(x_e)=\int\psi_{\mu 0}({\bf x}_\mu)(\frac{\alpha}{x_{\mu e}}-\frac{\alpha}{x_e})
\psi_{\mu 0}({\bf x}_\mu)d{\bf x}_\mu=-\frac{\alpha}{x_e}(1+W_\mu x_e)e^{-2W_\mu x_e}.
\end{equation}

The remaining integration over the coordinates of the electron can also
be performed analytically, giving in the following result:
\begin{equation}
\label{eq13}
\begin{array}{@{}l}\displaystyle
\Delta E_1^{(2)}(2S)=\int \psi_{e1}({\bf x}_e)\left(-\frac{\alpha}{x_e}\right)
(1+W_\mu x_e)e^{-2W_\mu x_e}d{\bf x}_e
\\[3mm]\displaystyle\hspace{22mm}
\times\int \psi_{e1}({\bf x'}_e)\left(-\frac{\alpha}{x'_e}\right)
(1+W_\mu x'_e)
e^{-2W_\mu x'_e}d{\bf x'}_e \;\widetilde{G}^e_{2S}(x_e,x_e')
\\[4mm]\displaystyle\hspace{20mm}
=-\frac{M_e\alpha^2}{8}\left[\frac{25}{16}a_1^3-
a_1^4\left(\frac{191}{32}+4\ln a_1\right)\right],
\end{array}
\end{equation}
where the final answer is presented in the form of an expansion in powers of the ratio 
of the effective masses of particles. Similarly, we can consider another contribution 
when the muon is in excited intermediate states $n\not=0$. This contribution 
can be represented as:
\begin{equation}
\label{eq14}
\begin{array}{@{}l}\displaystyle
\Delta E_2^{(2)}(2S)=\int \psi_{\mu 0}({\bf x}_\mu)\psi_{e1}({\bf x}_e)
\frac{\alpha}{|{\bf x}_\mu \!-\!{\bf x}_e|}
d{\bf x}_\mu d{\bf x}_e\sum_{n\not= 0}
\psi_{\mu n}({\bf x}_\mu)\psi_{\mu n}({\bf x'}_\mu)
\\[4mm]\displaystyle\hspace{25mm}
\times\> G^e({\bf x}_e,{\bf x'}_e,z)\frac{\alpha}{|{\bf x'}_\mu\!-\!{\bf x'}_e|}
\psi_{\mu 0}({\bf x'}_\mu)\psi_{e1}({\bf x'}_e)d{\bf x'}_\mu d{\bf x'}_e.
\end{array}
\end{equation}

The Coulomb Green's function of the electron standing here depends on the parameter
$ z=E_{\mu 0}-E_{\mu n} + E_{e1}$. In the leading order with respect
to the particle mass ratio, we choose $G^e ({\bf x}_e, {\bf x'}_e,z)$
in the form of the Green's function of a free electron:
\begin{equation}
\label{eq15}
G^e({\bf x}_e,{\bf x'}_e,z)=-\frac{M_e}{2\pi}\frac{1}{|{\bf x}_e\!-\!{\bf x'}_e|}
e^{-b |{\bf x}_e-{\bf x'}_e|}\>,
\qquad
b=\sqrt{2M_e(E_{\mu 0}\!-\!E_{\mu n}\!+\!E_{e1})}\>.
\end{equation}

Then the second part of the correction in the second order of the perturbation
theory takes the form:
\begin{equation}
\label{eq16}
\begin{array}{@{}l}\displaystyle
\Delta E_2^{(2)}(2S)=-\frac{M_e\alpha^2}{2\pi}
\int \psi_{\mu 0}({\bf x}_\mu)\psi_{e1}({\bf x}_e)\frac{1}{|{\bf x}_\mu\!-\!{\bf x}_e|}
d{\bf x}_\mu d{\bf x}_e
\sum_{n\not= 0}\psi_{\mu n}({\bf x}_\mu)\psi_{\mu n}({\bf x'}_\mu)
\\[3mm]\displaystyle\hspace{25mm}
\times\>\frac{1}{|{\bf x}_e\!-\!{\bf x'}_e|}e^{-b |{\bf x}_e-{\bf x'}_e|}
\frac{1}{|{\bf x'}_e\!-\!{\bf x'}_\mu|}
\psi_{\mu 0}({\bf x'}_\mu)\psi_{e1}({\bf x'}_e)d{\bf x'}_\mu d{\bf x'}_e.
\end{array}
\end{equation}

Replacing the value of the electron wave function with its value at zero
we first perform integration over ${\bf x'}_e$:
\begin{equation}
\label{eq17}
\begin{array}{@{}l}\displaystyle
I=\int d{\bf x'}_e\psi_{e1}({\bf x'}_e)\frac{1}{|{\bf x}_e\!-\!{\bf x'}_e||{\bf x'}_e\!-\!{\bf x'}_\mu|}
e^{b|{\bf x}_e\!-\!{\bf x'}_e|}
\\[3mm]\displaystyle\hspace{20mm}
=\frac{4\pi}{b^2}\,\psi_{e1}(0)
\>\frac{1}{|{\bf x}_e\!-\!{\bf x'}_\mu|}\>\left(1-e^{-b|{\bf x}_e-{\bf x'}_\mu|}\right)=
\\[3mm]\displaystyle\hspace{20mm}
=4\pi\,\psi_{e1}(0)\left[\frac{1}{b}-\frac{1}{2}|{\bf x}_e\!-\!{\bf x'}_\mu|+
\frac{b}{6}|{\bf x}_e\!-\!{\bf x'}_\mu|^2+\dots\right],
\end{array}
\end{equation}
where we also expanded the function in \eqref{eq17} in $b|{\bf x}_e-{\bf x'}_\mu| $. The term 
$1/b $ in square brackets vanishes due to the orthogonality of the muon wave 
functions. Leaving the leading term in square brackets, we will use the completeness 
relation in calculating the integrals over the coordinates of the particles:
\begin{equation}
\label{eq18}
\sum_{n\not =0}\psi_{\mu n}({\bf x}_3)\psi_{\mu n}^\ast({\bf x}_2)=
\delta({\bf x}_3-{\bf x}_2)-\psi_{\mu 0}({\bf x}_3)\psi_{\mu 0}^\ast({\bf x}_2).
\end{equation}

Direct calculation of integrals with functions from the right-hand side 
\eqref{eq16} gives the following results corresponding to two terms in \eqref{eq18} after 
expansion in $ a_1 $:
\begin{equation}
\label{eq19}
\Delta E^{(2)}_{21}(2S)=-16M_e\alpha^2,~~~\Delta E^{(2)}_{22}(2S)=
M_e\alpha^2\bigl(16+\frac{105}{128}a_1^3-\frac{481}{512}a_1^4\bigr).
\end{equation}

Let us now calculate the second-order correction of the perturbation theory for 
the $2P$ state. Let us first choose a muon in an intermediate state with $n =0 $. 
Such a contribution will be determined by the following integral:
\begin{equation}
\label{eq20}
\Delta E_1^{(2)}(2P)=\int \psi_{e1}({\bf x}_e)\left(-\frac{\alpha}{x_e}\right)
(1+W_\mu x_e)e^{-2W_\mu x_e}d{\bf x}_e
\int \psi_{e1}({\bf x'}_e)\left(-\frac{\alpha}{x'_e}\right)\times
\end{equation}
\begin{displaymath}
(1+W_\mu x'_e)e^{-2W_\mu x'_e}d{\bf x'}_e \tilde G^e_{2P}(x_e,x'_e).
\end{displaymath}

The reduced Coulomb Green's function of the electron for the excited state $2P $
has the following form:
\begin{equation}
\label{eq21}
\begin{array}{@{}l}\displaystyle
\widetilde{G}(2P)=-\frac{Z\alpha\mu^2}{36x^2_1x^2_2}\>e^{-\frac{x_1+x_2}{2}}\>
\frac{3}{4\pi}\frac{({\bf x}_1{\bf x}_2)}{x_1x_2}\>g_{2P}(x_1,x_2),
\\[4mm]
g_{2P}(x_1,x_2)=24x^3_<+36x^3_<x_>+36x^3_<x^2_>+24x^3_>+36x_<x^3_>+36x^2_<x^3_>
\\[3mm]\hspace{25mm}
+49x^3_<x^3_>-3x^4_<x^3_>-12e^{x_<}(2+x_<+x_<^2)x^3_>-3x^3_<x^4_>
\\[3mm]\hspace{25mm}
+12x_<^3x_>^3\left[-2C+\mathrm{Ei}(x_<)-\ln(x_<)-\ln(x_>)\right].
\end{array}
\end{equation}

As in the case of the $2S$ state, all coordinate integrals can be calculated 
analytically. Expanding the final result in powers of $a_1=W_e /W_\mu $ 
with precision $O(a_1^6)$, we get:
\begin{equation}
\label{eq22}
\Delta E^{(2)}(2P)=-M_e\alpha^2\Bigl[\frac{7}{2048}a_1^5-\frac{9}{8192}a_1^6\Bigr].
\end{equation}
As follows from the expansions \eqref{eq7}, \eqref{eq22}, the order of the correction 
for the $2P$ state increases in comparison with the $2S$ state of the electron, 
and its magnitude decreases significantly. Let us study 
the second contribution for the $2P$ state connected with muonic excitations
which initially can be written in the form:
\begin{equation}
\label{eq22a}
\Delta E_2^{(2)}(2P)=-\frac{M_e\alpha^2}{8\pi^2}\int\psi_{\mu 0}({\bf x}_\mu)
\psi_{e2P}({\bf x}_e)\frac{d{\bf x}_e}{|{\bf x}_\mu-{\bf x}_e|}d{\bf x}_\mu
\sum_{n\not=0}\psi_{\mu n}({\bf x}_\mu)\psi_{\mu n}({\bf x}'_\mu)
\frac{d{\bf x}'_e}{|{\bf x}'_\mu-{\bf x}'_e|}
\end{equation}
\begin{displaymath}
e^{-b|{\bf x}_e-{\bf x}'_e|}\frac{1}{|{\bf x}_e-{\bf x}'_e|}\psi_{\mu 0}({\bf x}'_{\mu})
\psi_{e2P}({\bf x}'_e).
\end{displaymath}
After the variables shift ${\bf x}_e\to {\bf x}_e+{\bf x}_\mu$, ${\bf x}'_e\to {\bf x}'_e+{\bf x}'_\mu$
we can neglect ${\bf x}_\mu$, ${\bf x}'_\mu$ in electron wave functions and present the
leading order contribution in the mass ratio $M_e/M_\mu$ in \eqref{eq22a} as follows:
\begin{equation}
\label{eq22b}
\Delta E_2^{(2)}(2P)=-\frac{M_e\alpha^2}{192\pi^2}W_e^5\int \frac{d{\bf x}_e}{x_e}\frac{d{\bf x}'_e}{x'_e}
e^{-\frac{1}{2}W_ex_e}e^{-\frac{1}{2}W_ex'_e}e^{-b|{\bf x}_e-{\bf x}'_e|}\frac{1}{|{\bf x}_e-{\bf x}'_e|}\times
\end{equation}
\begin{displaymath}
\sum_{n\not=0}\psi_{\mu n}({\bf x}_\mu){\bf x}_\mu\psi_{\mu 0}({\bf x}_\mu)
\psi_{\mu n}({\bf x}'_\mu){\bf x}'_\mu\psi_{\mu 0}({\bf x}'_\mu).
\end{displaymath}
Then all integrals over the electron coordinates can be calculated analytically
giving the result:
\begin{equation}
\label{eq22c}
\Delta E_2^{(2)}(2P)=-\frac{M_e\alpha^2(Z-1)}{36Z}\frac{W_e^3}{W_\mu^3}(S^d+S^c)=
\begin{cases}
{}_3^{7}\mbox{Li}\,\mu e:~-0.014~\mbox{GHz}.\\
{}_4^9\mbox{Be}\,\mu e:~-0.023~\mbox{GHz},\\
{}_5^{11}\mbox{B}\,\mu e:~-0.029~\mbox{GHz},\\
\end{cases}
\end{equation}
\begin{displaymath}
S^d=\sum_{n=2}^\infty\frac{n^2}{n^2-1}|<\psi_{\mu 0}|W_\mu{\bf x}_\mu|\psi_{\mu n}>|^2=
\sum_{n=2}^\infty\frac{2^8n^9(n-1)^{2n-6}}{(n+1)^{2n+6}}=2.747443...,
\end{displaymath}
\begin{displaymath}
S^c=\int_0^\infty kdk\frac{2^8}{(1-e^{-\frac{2\pi}{k}})}\frac{1}{(k^2+1)^6}|\left(\frac{1+Ik}{1-Ik}\right)^{I/k}|^2=
0.627556...
\end{displaymath}

The results obtained show that the expansion parameter, which 
works in the framework of perturbation theory in $ \Delta H$ \eqref{eq2}, is the 
particle mass ratio $ M_e/M_\mu \approx m_e /m_\mu$.

Let us conclude this section by calculating the interaction correction 
$\Delta H$ in the second order of the perturbation theory for the electronic state 
$ 1S$:
\begin{equation}
\label{eqq9}
\Delta E^{(2)}(1S)=\int \psi_{\mu 0}({\bf x}_\mu)\psi_{e0}({\bf x}_e)(\frac{\alpha}{x_{\mu e}}-\frac{\alpha}{x_e})
\sum_{n,n'}\frac{\psi_{\mu n}({\bf x}_\mu)\psi_{en'}({\bf x}_e)\psi_{\mu n}({\bf x'}_\mu)
\psi_{en'}({\bf x'}_e)}{E_{\mu 0}+E_{e0}-E_{\mu n}-E_{en'}}\times
\end{equation}
\begin{displaymath}
\psi_{\mu 0}({\bf x'}_\mu)\psi_{e0}({\bf x'}_e)
(\frac{\alpha}{x'_{\mu e}}-\frac{\alpha}{x'_e})d{\bf x}_\mu 
d{\bf x'}_\mu d{\bf x}_e d{\bf x'}_e.
\end{displaymath}
Let us first extract, as before, the contribution of the muon in the ground 
state $ n = 0 $:
\begin{equation}
\label{eqq10}
\Delta E_1^{(2)}(1S)=\int \psi_{\mu 0}({\bf x}_\mu)\psi_{e0}({\bf x}_e)(\frac{\alpha}{x_{\mu e}}-\frac{\alpha}{x_e})\psi_{\mu 0}({\bf x}_\mu)\psi_{\mu 0}({\bf x'}_\mu)
\sum_{n'}\frac{\psi_{en'}({\bf x}_e)\psi_{en'}({\bf x'}_e)}{E_{e0}-E_{en'}}\times
\end{equation}
\begin{displaymath}
\psi_{\mu 0}({\bf x'}_\mu)\psi_{e0}({\bf x'}_e)
(\frac{\alpha}{x'_{\mu e}}-\frac{\alpha}{x'_e})d{\bf x}_\mu d{\bf x'}_\mu d{\bf x}_e 
d{\bf x'}_e.
\end{displaymath}
The reduced Coulomb Green's function of the electron for the ground state 
$ \tilde G^e_{1S} $ has the following form:
\begin{equation}
\label{eq11}
\tilde G^e_{1S}(r_1,r_2)=-\frac{(Z-1)\alpha M_e^2}{\pi}e^{-(x_1+x_2)}
g_{1S}(x_1,x_2),
\end{equation}
\begin{displaymath}
g_{1S}(x_1,x_2)=\frac{1}{2x_>}-\ln 2x_>-\ln 2x_<+Ei(2x_<)+\frac{7}{2}-2\gamma-
(x_1+x_2)+\frac{1}{2x_<}(1-e^{2x_<}).
\end{displaymath}
As in the case of \eqref{eq14}, all integrals are calculated analytically, 
and the result of the calculation is:
\begin{equation}
\label{eq11a}
\Delta E^{(2)}_1(1S)=-M_e\alpha^2\left[\frac{25}{16}a_1^3+a_1^4\left(
\frac{1}{32}-4\ln 2-4\ln a_1\right)+a_1^5\left(
-\frac{565}{32}+20\ln 2+20\ln a_1\right)\right],
\end{equation}
which we presented in the form of expansion in terms of $a_1 $. Since the expansion 
starts from the third power of $ a_1$, this contribution to the $ (2S-1S) $ interval 
is numerically small compared to the contribution of the leading order.

When calculating the contribution of muonic intermediate states with $ n\not= 0$, 
we replace the Green's function of the electron by the free Green's function 
\eqref{eq15} and use the relation
completeness, as in \eqref{eq18}. Then the second part of the correction will 
be determined by the expression:
\begin{equation}
\label{poln}
\begin{array}{@{}l}\displaystyle
\Delta E_2^{(2)}(1S)=M_e\alpha^2\psi_{e0}(0)\int \psi_{\mu 0}({\bf x}_2\psi_{e0}({\bf x}_1)
\frac{|{\bf x}_1-{\bf y}_2|}{|{\bf x}_1-{\bf x}_2|}d{\bf x}_1d{\bf x}_2
\\[4mm]\displaystyle\hspace{30mm}
\times\>\psi_{\mu 0}({\bf y}_2)d{\bf y}_2\left[\delta({\bf x}_2-{\bf y}_2)-
\psi_{\mu 0}({\bf x}_2)\psi_{\mu 0}^\ast({\bf y}_2)\right]
\end{array}
\end{equation}
The $ \delta $ -function term contributes
$\Delta E_{21}^{(2)}(1S)=8M_e\alpha^2 $, and the second part in the
completeness condition reduces to the integral
\begin{equation}
\label{poln1}
\begin{array}{@{}l}\displaystyle
\Delta E^{(2)}_{22}(1S)=-2M_e\alpha^2a_1^3\int_0^\infty dx e^{-4x(1+a_1/4)}
(-1\!-\!x\!+\!e^{2x})(-2\!-\!x\!+\!2e^{2x}(1\!+\!x^2)]=
\\[4mm]\displaystyle\hspace{21mm}
=-2M_e\alpha^2\left(4+2a_1^2-\frac{105}{32}a_1^3\right).
\end{array}
\end{equation}
As a result, the second part of the correction in the second order of the
perturbation theory for the $ 1S$ state takes the form:
\begin{equation}
\label{poln2}
\Delta E^{(2)}_2(1S)=\Delta E^{(2)}_{21}(1S)+\Delta E^{(2)}_{22}(1S)=
-2M_e\alpha^2\left[2a_1^2-\frac{105}{32}a_1^3\right].
\end{equation}

\section{Vacuum polarization, nuclear structure and recoil corrections}

The effect of electronic polarization of the vacuum leads to the appearance of a 
number of additional terms in the Hamiltonian of the interaction of three particles, 
which are determined by the following expressions:
\begin{equation}
\label{vp1}
\Delta V_{vp}^{eN}(x_e)=\frac{\alpha}{3\pi}\int_1^\infty
\rho(\xi)\left(-\frac{2\alpha}{x_e}\right) e^{-2m_e\xi
x_e}d\xi,~~~\rho(\xi)=\frac{\sqrt{\xi^2-1}(2\xi^2+1)}{\xi^4},
\end{equation}
\begin{equation}
\label{vp2}
\Delta V_{vp}^{\mu N}(x_\mu)=\frac{\alpha}{3\pi}\int_1^\infty
\rho(\xi)\left(-\frac{2\alpha}{x_\mu}\right) e^{-2m_e\xi x_\mu}d\xi,
\end{equation}
\begin{equation}
\label{vp3}
\Delta V_{vp}^{e\mu}(|{\bf x}_e-{\bf x}_\mu|)=\frac{\alpha}{3\pi}\int_1^\infty
\rho(\xi)\frac{\alpha}{x_{e\mu}} e^{-2m_e\xi x_{e\mu}}d\xi,
\end{equation}

When studying the energy levels of light muonic atoms, it was found that the contribution 
of the electronic polarization of the vacuum is the main one both for the Lamb shift 
$ (2P-2S)$ and for the $(2S-1S)$ interval. Therefore, we will take into account 
the correction for vacuum polarization to the Coulomb potential \eqref{vp2}. 
For any electron states $ 1S$, $ 2S $, $ 2P$, this correction in the energy spectrum 
is the same; therefore, for the intervals $ (2P-2S) $, $ (2S-1S) $, the contribution 
is zero. In the second order of perturbation theory, the vacuum polarization 
contribution \eqref{vp2} is determined by the following integral expression:
\begin{equation}
\label{q1}
\Delta E^{(2)~\mu N}_{vp}=\frac{2\alpha}{3\pi}\int \psi_{\mu 0}({\bf x}_\mu)
\psi_{e~nL}({\bf x}_e)
\left(-\frac{Z\alpha}{x_\mu}\right)
\sum_{n,n'}\frac{\psi_{\mu n}({\bf x}_\mu)\psi_{en'}({\bf x}_e)\psi_{\mu n}({\bf x'}_\mu)
\psi_{en'}({\bf x'}_e)}{E_{\mu 0}+E_{e~nL}-E_{\mu n}-E_{en'}}\times
\end{equation}
\begin{displaymath}
\rho(\xi)e^{-2m_e\xi x_\mu}
\psi_{\mu 0}({\bf x'}_\mu)\psi_{e~nL}({\bf x'}_e)
(\frac{\alpha}{x'_{\mu e}}-\frac{\alpha}{x'_e})d{\bf x}_\mu 
d{\bf x'}_\mu d{\bf x}_e d{\bf x'}_e.
\end{displaymath}
Taking into account the orthogonality of the electron wave functions in the initial 
and intermediate states, it is necessary to set $n'=nL $ ($1S $, $ 2S$, $2P $). 
Then the difference in the contribution of the muon-nuclear polarization of the vacuum 
will be connected with the remaining integral over the coordinates of the electron 
in this matrix element. For the states $1S $, $2S $, $2P $ these integrals have 
the form:
\begin{equation}
\label{j1}
J_1=\int |\psi_{1S}^{e}({\bf x'}_e)|^2\left(
\frac{\alpha}{x'_{\mu e}}-\frac{\alpha}{x'_e}\right)=
2\alpha W_e\frac{e^{-W_ex_\mu'}}{W_e x'_\mu}\left[-W_ex'_\mu \cosh(W_ex_\mu')+
\sinh(W_ex'_\mu)
\right],
\end{equation}
\begin{equation}
\label{j2}
J_2=\int |\psi_{2S}^{e}({\bf x'}_e)|^2\left(
\frac{\alpha}{x'_{\mu e}}-\frac{\alpha}{x'_e}\right)=
\frac{\alpha W_e}{4}\frac{1}{2W_e x'_\mu}\Bigl[
8-2W_ex'_\mu+e^{-W_ex'_\mu}(-8-W_ex'_\mu\times
\end{equation}
\begin{displaymath}
(6+W_ex'_\mu(2+W_ex'_\mu)))
\Bigr],
\end{displaymath}
\begin{equation}
\label{j3}
J_3=\int |\psi_{2P}^{e}({\bf x'}_e)|^2\left(
\frac{\alpha}{x'_{\mu e}}-\frac{\alpha}{x'_e}\right)=
\frac{\alpha W_e}{48}\frac{1}{W_e x'_\mu}\Bigl[-12W_ex'_\mu-36W_ex'_\mu e^{-W_ex'_\mu}+
\end{equation}
\begin{displaymath}
48(1-e^{-W_ex'_\mu})-12(W_ex'_\mu)^2e^{-W_ex'_\mu}-2(W_ex'_\mu)^3e^{-W_ex'_\mu}
\Bigr].
\end{displaymath}
Given the explicit expressions for $J_i$, it is possible to perform analytical 
integration over the particle coordinates in \eqref{q1}. Expanding in the parameter 
$a_1$ and integrating over the remaining spectral parameter, we obtain the contribution 
in the following form:
\begin{equation}
\label{q2}
\Delta E^{(2)~\mu N}_{vp}(1S)=\frac{64\alpha^2W_\mu}{3\pi}\int\rho(\xi)d\xi\int_0^\infty
x_\mu dx_\mu\int_0^\infty x'_\mu dx'_\mu g_{1S}(x_\mu,x'_\mu)e^{-2x_\mu(1+a_2\xi)}
e^{-2x'_\mu(1+a_1/2)}\times
\end{equation}
\begin{displaymath}
\left[-a_1x'_\mu \cosh (a_1x'_\mu)+\sinh (a_1x'_\mu)\right]=
\frac{\alpha^2W_\mu a_1^3}{432\pi(1-a_2^2)^{9/2}}\Bigl\{
-8(1-a_2^2)^{3/2}(-152+978a_2^2-
\end{displaymath}
\begin{displaymath}
2763a_2^4+2924a_2^6-1032a_2^8-516\pi a_2^3+1548\pi a_2^5-1548\pi a_2^7+516\pi a_2^9)+
24(48-216a_2^2+
\end{displaymath}
\begin{displaymath}
948a_2^4-2285a_2^6+2709a_2^8-1548a_2^{10}+344a_2^{12})\ln\frac{a_2}{1-\sqrt{1-a_2^2}}
\Bigr\}.
\end{displaymath}
\begin{equation}
\label{q3}
\Delta E^{(2)~\mu N}_{vp}(2S)=
\frac{\alpha^2W_\mu a_1^3}{3456\pi(1-a_2^2)^{4}}\Bigl\{-\bigl[
1216+a_2^2(-9040+a_2(29928a_2-45496a_2^3+31648a_2^5-8256a_2^7+
\end{equation}
\begin{displaymath}
4128(1-a_2^2)\pi\bigr]+24\sqrt{1-a_2^2}\bigl(-48+168a_2^2-780a_2^4+1505a_2^6-
1204a_2^8+344a_2^{10}\bigr)\ln\frac{a_2}{1+\sqrt{1-a_2^2}}\Bigr\},
\end{displaymath}
\begin{equation}
\label{q4}
\Delta E^{(2)~\mu N}_{vp}(2P)=
\frac{\alpha^2W_\mu a_1^5}{230400\pi(1-a_2^2)^{11/2}}\Bigl\{
\sqrt{1-a_2^2}\bigl(32804-290020a_2^2+1252145a_2^4-2539811a_2^6+
\end{equation}
\begin{displaymath}
2636152a_2^8-1368800a_2^{10}+283200a_2^{12}+141600\pi a_2^3-708000\pi a_2^5+
14116000\pi a_2^7-1416000\pi a_2^9+
\end{displaymath}
\begin{displaymath}
708000\pi a_2^{11}-141600\pi a_2^{13}\bigr)+15\bigl(-1920+10560a_2^2-59010a_2^4+
174877a_2^6-273565a_2^8+
\end{displaymath}
\begin{displaymath}
233640a_2^{10}-103840a_2^{12}+18880a_2^{14}\Bigr]\ln\frac{a_2}
{1-\sqrt{1-a_2^2}}\Bigr\},
\end{displaymath}
where the parameter $a_2=m_e/W_\mu$.
We have presented here only the first terms of the expansions in the parameter 
$ a_1$, which give the numerical values of the contributions with high accuracy.

Let us consider the calculation of other corrections for vacuum polarization, 
which are determined by the potentials \eqref{vp1}, \eqref{vp3} and appear already 
in the first order of the perturbation theory. In the electron-nuclear interaction, 
the correction for vacuum polarization for $S$ states is determined by the 
following expression:
\begin{equation}
\label{q44}
\Delta E^{(1)~eN}_{vp}(nS)=-\frac{4\alpha(Z\alpha)m_e}{15\pi}
\left(\frac{W_e}{m_e}\right)^3\delta_{l0}.
\end{equation}
To estimate the value of the energy interval $(2S-1S)$, this correction is not very significant, since it has values of tens of GHz. It is more important for the total 
value of the Lamb shift $(2P-2S)$. In the case of the $2P$ state, the correction 
for vacuum polarization has the form
\begin{equation}
\label{q44a}
\Delta E^{(1)~eN}_{vp}(2P)=-\frac{\alpha(Z\alpha)m_e}{560\pi}
\left(\frac{W_e}{m_e}\right)^5
\end{equation}
and is very small. Therefore, the value of this correction to the shift $(2P-2S)$ is determined by \eqref{q44}:
\begin{equation}
\label{q45}
\Delta E^{(1)~eN}_{vp}(2P-2S)=
\begin{cases}
{}_3^{7}\mbox{Li}\,\mu e:~0.650~\mbox{GHz}.\\
{}_4^9\mbox{Be}\,\mu e:~2.929~\mbox{GHz},\\
{}_5^{10}\mbox{B}\,\mu e:~8.680~\mbox{GHz},\\
\end{cases}
\end{equation}

The correction of the muon-electron vacuum polarization (potential \eqref{vp3}) 
is also important for the $ (2P-2S)$ shift. Taking into account the expansion terms 
$a_1^2$ and $ a_1^3$, it takes the form:
\begin{equation}
\label{q46}
\Delta E^{(1)~e\mu}_{vp}(2P-2S)=-\frac{\alpha^2 W_e}{3\pi}a_1^2\Bigl\{
\frac{2}{5a_2^2}+\frac{a_1}{480a_2^4\sqrt{4-a_2^2}}\Bigl[
\sqrt{4-a_2^2}\Bigl(-620a_2^4-120a_2^6-150\pi a_2-
\end{equation}
\begin{displaymath}
270\pi a_2^3+135\pi a_2^5+30\pi a_2^7\Bigr)+30a_2^4(-40+5a_2^2+2a_2^4)
\ln\frac{2-\sqrt{4-a_2^2}}{a_2}\Bigl]\Bigl\}=
\begin{cases}
_3^7Li\mu e:-0.212~GHz,\\
_4^9Be\mu e:-0.711~GHz,\\
_5^{11}B\mu e:-1.670~GHz.
\end{cases}
\end{displaymath}

Another correction in the energy spectrum that improves the accuracy of the results 
is the correction for the structure of the nucleus. In two-particle muonic atoms, 
the correction for the structure of the nucleus is significant, since the muon 
is closer to the nucleus than the electron. The interaction potentials of a muon 
and a nucleus, an electron and a nucleus, necessary for calculating the correction, 
are determined by the following formulas:
\begin{equation}
\label{q5}
\Delta V_{str}^{\mu N}({\bf x}_\mu)=\frac{2}{3}\pi(Z\alpha)r^2_N\delta({\bf x}_\mu),~~~
\Delta V_{str}^{e N}({\bf x}_e)=\frac{2}{3}\pi(Z\alpha)r^2_N\delta({\bf x}_e),
\end{equation}
where $r_N$ is the nucleus charge radius.
The potential $\Delta V_{str}^{\mu N}({\bf x}_\mu)$ does not contribute to the 
shift $(2S-1S)$, $(2P-2S)$ in the first order of the perturbation theory , 
since it depends only on the muon coordinate. 
The $\Delta V_{str}^{\mu N}({\bf x}_\mu)$ contribution in the second 
order of the perturbation theory is expressed in terms of the reduced Coulomb 
Green's function of the muon $1S $ state with one zero argument, which has the 
following form:
\begin{equation}
\label{gr1}
\tilde G^\mu_{1S}({\bf r})=\frac{Z\alpha M_\mu^2}{4\pi}\frac{e^{-x}}{x}g_{1S}(x),~
g_{1S}(x)=\left[4x(\ln 2x+C)+4x^2-10x-2\right].
\end{equation}
The different kind of correction for the electron states $1S $, $ 2S$, $ 2P$ 
is determined by the same functions $J_i$ \eqref{j1}-\eqref{j3} as for 
vacuum polarization. The initial integral expression for this correction is presented 
as follows:
\begin{equation}
\label{q6}
\Delta E^{(2)~\mu N}_{str}(nL)=\frac{4}{3}(Z\alpha)r_N^2W_\mu^3\int 
\tilde G^\mu_{1S}({\bf x}_\mu)J_i({\bf x}_\mu)e^{-W_\mu x_\mu}d{\bf x}_\mu.
\end{equation}
For all electronic states, coordinate integration is performed analytically, and 
the results can be presented as:
\begin{equation}
\label{q7}
\Delta E^{(2)~\mu N}_{str}(1S)=\frac{8}{3}\alpha(Z\alpha)r_N^2W_\mu^2M_\mu a_1^3
\left(-\frac{11}{6}+\frac{155}{24}a_1\right),
\end{equation}
\begin{equation}
\label{q8}
\Delta E^{(2),\mu N}_{str}(2S)=\frac{1}{3}\alpha(Z\alpha)r_N^2W_\mu^2M_\mu a_1^3
\left(-\frac{11}{6}+\frac{155}{24}a_1\right),
\end{equation}
\begin{equation}
\label{q9}
\Delta E^{(2)~\mu N}_{str}(2P)=-\frac{1}{2880}\alpha(Z\alpha)r_N^2W_\mu^2M_\mu a_1^5
\left(591-1659a_1\right).
\end{equation}
The high degree of $ a_1 $ in \eqref{q9} for the $ 2P $ state makes this correction 
in the Lamb shift negligible. The second potential from \eqref{q5} gives
a known contribution already in the first order of the perturbation theory:
\begin{equation}
\label{q10}
\Delta E^{(1),e N}_{str}(nS)=\frac{2}{3n^3}(Z\alpha)^4r_N^2M_e^3\delta_{l0}.
\end{equation}
The numerical values of the corrections \eqref{q8}, \eqref{q10} are important 
to refine the results on the Lamb shift, since they are tenths of a GHz.

Nuclear recoil contribution is determined by the Hamiltonian $\Delta H_{rec}$.
In the first order perturbation theory recoil correction is equal zero 
due to the vanishing of the integral over the angular variables.
For the same reason, the state of the muon with $n = 0$ in the second order of
perturbation theory does not contribute. The muon contribution with $n\not =0$
can be written in the second order PT as
\begin{equation}
\label{q99}
\Delta E^{(2)}_{rec}(2S)=-\frac{M_eW_\mu^2W_e^2}{8\pi m_N^2}\int\psi_{\mu 0}({\bf x}_\mu)
\frac{\partial}{\partial x'_e}\psi_{e2S}({\bf x}_e)\frac{{\bf x}_\mu{\bf x}_e}{x_\mu x_e}d{\bf x}_\mu d{\bf x}_e
\sum_{n\not=0}\psi_{\mu n}({\bf x}_\mu)\psi^\ast_{\mu n}({\bf x'}_\mu)\times
\end{equation}
\begin{displaymath}
\frac{e^{-b|{\bf x}_e-{\bf x'}_e|}}{|{\bf x}_e-{\bf x'}_e|}
\frac{{\bf x'}_\mu{\bf x'_e}}{x'_\mu x'_e}\psi_{\mu 0}({\bf x'}_\mu)
\frac{\partial}{\partial x'_e}\psi_{e2S}({\bf x'}_e)
d{\bf x'}_\mu d{\bf x'}_e.
\end{displaymath}
In this expression, the following integral symmetric tensor in the coordinates of the electron 
can be distinguished:
\begin{equation}
\label{q999}
J^{ij}=\int d{\bf x}_e d{\bf x'}_e \frac{{x_e}^i {x'_e}^j}{x_ex'_e}
\frac{e^{-b|{\bf x}_e-{\bf x'}_e|}}{|{\bf x}_e-{\bf x'}_e|}\frac{\partial}{\partial x'_e}\psi_{e2S}({\bf x}_e)
\frac{\partial}{\partial x'_e}\psi_{e2S}({\bf x'}_e)=\delta^{ij}A,
\end{equation}
where the value of the integral is found after convolution with the tensor $\delta^{ij}$:
\begin{equation}
\label{nn0}
A=\frac{\pi}{3W_e^2}(4b_1^2-13b_1^4),~~~b_1^2=\frac{M_e}{M_\mu}\frac{(Z-1)^2}{Z^2}\frac{n^2}{(n^2-1)}.
\end{equation}
A similar calculation for the $2P$ state gives a factor $(4b_1^2-\frac{7}{3}b_1^4)$ in the corresponding 
integral for the $2P$ state. In the difference, the first terms cancel out, and the final result has the form:
\begin{equation}
\label{nn1}
\Delta E^{(2)}_{rec}(2P-2S)=-\frac{16}{9}\frac{M_e\alpha^2M_e^2(Z-1)^4}{Z^2m_N^2}\int d{\bf x}_\mu\int 
d{\bf x'}_\mu\sum_{n\not=0}\psi^\ast_{\mu 0}({\bf x}_\mu)\frac{{\bf x}_\mu}{x_\mu}
\psi_{\mu n}({\bf x}_\mu)\times
\end{equation}
\begin{displaymath}
\psi^\ast_{\mu 0}({\bf x'}_\mu)\frac{{\bf x'}_\mu}{x'_\mu}\psi_{\mu n}({\bf x'}_\mu)\frac{n^4}{(n^2-1)^2}
\end{displaymath}

The square of the radial integral for the transition $1S\to nP$ has the form:
\begin{equation}
\label{nn1a}
(I_{1S}^{nP})^2=\frac{4}{n(n^2-1)}\left[1-\frac{(5n^2-1)}{(n^2-1)}\frac{(n-1)^n}{(n+1)^n}\right]^2,~~~
I_{1S}^{nP}=\int_0^\infty R_{10}(r)R_{n1}(r)rdr.
\end{equation}

Then the contributions of discrete and continuous spectrum in \eqref{nn1} are determined by 
following expressions:
\begin{equation}
\label{nn1b}
C^d=\sum_{n=2}^\infty \frac{4n^3}{(n^2-1)^3}\left[1-(5n^2-1)\frac{(n-1)^{n-1}}{(n+1)^{n+1}}\right]=0.474899...,
\end{equation}
\begin{equation}
\label{nn1c}
C^c=\int_0^\infty\frac{4kdk}{(k^2+1)^3(1-e^{-\frac{2\pi}{k}})}\left[1-\frac{(5+k^2)}{(1+k^2)}\left(\frac{1+Ik}{1-Ik}\right)^{\frac{I}{k}}\right]^2=0.129105...
\end{equation}

As a result, the total value of the correction \eqref{nn1} in the Lamb shift is
\begin{equation}
\label{nn1d}
\Delta E^{(2)}_{rec}(2P-2S)=-\frac{16}{27}\frac{M_e\alpha^2M_e^2(Z-1)^4}{Z^2m_N^2}(C^d+C^c)=
\begin{cases}
_3^7Li\mu e:-0.026~GHz,\\
_4^9Be\mu e:-0.044~GHz,\\
_5^{11}B\mu e:-0.060~GHz.
\end{cases}
\end{equation}

\section{Numerical results}

The total analytical contribution of the perturbation operators $ \Delta H $ and $\Delta H_{rec}$
obtained in the first and second orders of the perturbation theory to the 
Lamb shift $(2P-2S)$ is:
\begin{equation}
\label{eq23}
\Delta E(2P-2S)=M_e\alpha^2\Bigl[\frac{Z-1}{4}a_1^2-\frac{5Z}{8}a_1^3-
\frac{(Z-1)}{36Z}a_1^3(S^d+S^c)+
\Bigl(\frac{15Z}{16}-\frac{381}{512}\bigr)a_1^4-
\end{equation}
\begin{displaymath}
-\frac{16}{27}\frac{M_e^2(Z-1)^4}{Z^2m_N^2}(C^d+C^c)+
\bigl(\frac{7Z}{64}-
\frac{231}{2048}\bigr)a_1^5+\frac{9}{8192}a_1^6-\frac{1}{2}a_1^4\ln a_1\Bigr],
\end{displaymath}
where terms of the order $ O(a_1^5) $ and $ O (a_1^6) $ are taken only for the 
$2P$ state. Similarly, the energy interval $(2S-1S) $ obtained with allowance for 
recoil effects according to the perturbation theory with the Hamiltonian $ \Delta H $ 
is:
\begin{equation}
\label{eq23a}
\Delta E(2S-1S)=M_e\alpha^2\Bigl[
\frac{(Z-1)^2}{2}+\frac{7}{4}(Z-1)a_1^2+\bigl(5-\frac{25}{8}Z\bigr)a_1^3+
\bigl(\frac{513}{64}Z-\frac{4203}{512}\bigr)a_1^4+\frac{1}{2}a_1^4\ln a_1
\Bigr].
\end{equation}

Using the values of the fundamental parameters of the theory, one can obtain from 
the formulas \eqref{eq23}, \eqref{eq23a} numerical estimates of the electron Lamb 
shift and the $ (2S-1S) $ interval for muonic lithium ions ($_3^7Li $), 
beryllium ($_4^9Be $) and boron ($_5^{11} B $), which are presented in 
Tables~\ref{tb1}, \ref{tb2}.

\begin{table}[htbp]
\caption{Electronic energy interval $(2S-1S)$ in GHz.} 
\label{tb1}
\medskip
\begin{tabular}{|c|c|c|c|} \hline 
Contribution & $(\mu e _3^7Li)$  &$(\mu e _4^9Be)$  & $(\mu e _5^{11}B)$ \\ \hline 
Eq.\eqref{eq23a}& $9.868999\cdot 10^6$&$22.205543\cdot 10^6$ &$39.476833\cdot 10^6$ \\ \hline
Vacuum polarization correction   &  2.140     & 13.418    & 45.433   \\     \hline
Nuclear structure correction     &  -0.188   &  -0.527    & -1.286   \\     \hline 
Relativistic correction     &  481.832   &  2439.636    & 7711.943   \\     \hline
QED correction        &  -95.367   &  -420.460    & -1189.068   \\     \hline
Summary contribution  &$9.8690\cdot 10^6$ & $22.2055\cdot 10^6$ & $39.4768\cdot 10^6$ \\     \hline
\end{tabular}
\end{table}

\begin{table}[htbp]
\caption{Electronic Lamb shift $(2P-2S)$ in GHz.} 
\label{tb2}
\medskip
\begin{tabular}{|c|c|c|c|} \hline 
Contribution & $(\mu e _3^7Li)$  &$(\mu e _4^9Be)$  & $(\mu e _5^{11}B)$ \\ \hline 
Eq.\eqref{eq23} &  34.837 &65.684 & 99.214  \\     \hline
Vacuum polarization correction        &0.306   & 1.916       &  6.491   \\     \hline
Nuclear structure correction    &  -0.027  & -0.075   &  -0.184      \\     \hline 
Relativistic correction      &  0    &    0       &      0    \\     \hline
QED correction      & -14.257     & -63.273     & -180.004    \\     \hline
Summary contribution      & 20.859  & 4.252    &  -74.483  \\     \hline
\end{tabular}
\end{table}

The calculation of the energy intervals $(2S-1S)$ and $(2P-2S)$ is also performed 
by us within the framework of the variational method formulated in 
\cite{korobov,korobov1} with very high accuracy (see section V). Our analytical results coincide 
with numerical calculations of $ (2S-1S) $ in the variational approach with an accuracy 
of 0.0001 GHz. In the case of the Lamb shift, the difference between the results from 
Table~\ref{tb2} (see line 1) from the variational calculations turns out to be 
more significant. Here are the results of the variational method with an accuracy 
of 0.001 GHz:
\begin{equation}
\label{eq24}
\begin{array}{@{}l}
\Delta E_{\rm Li}(2P-2S)=36.568~\mbox{GHz},~
\\[3mm]
\Delta E_{\rm Be}(2P-2S)=68.015~\mbox{GHz},~
\\[3mm]
\Delta E_{\rm B}(2P-2S)=101.947~\mbox{GHz}.
\end{array}
\end{equation}

This difference between \eqref{eq24} and the results in line 1 of Table~\ref{tb2}
is due to the approximation that we use in analytical calculations 
in the second order of the perturbation theory. Expanding in the parameter $ b$ 
\eqref{eq17}, we took into account the leading order correction in $M_e /M_\mu $, 
and therefore neglected the contributions of the order $O(\sqrt{M_e/M_\mu}) $ 
with respect to accounted contribution.

The corrections for the nuclear structure and vacuum polarization calculated in Section~III 
are included in Tables~\ref{tb1}, \ref{tb2} in separate lines. They are important 
for refining the Lamb shift $ (2P-2S)$. There are two more important corrections that 
must be taken into account when obtaining the total numerical value of the energy interval:
the relativistic correction and the QED correction, which is the main one for obtaining 
the Lamb shift in the hydrogen atom. An analytical expression is known for it, which 
we represent in the form \cite{egs}:
\begin{equation}
\label{qed}
\Delta E_{QED}(nS)=\frac{\alpha((Z-1)\alpha)^4}{\pi n^3}\frac{M_e^3}{m_e^2}
\Bigl[\frac{4}{3}\ln\frac{m_e}{M_e(Z-1)^2\alpha^2}-\frac{4}{3}\ln k_0(nS)+\frac{10}{9}
\Bigr],
\end{equation}
\begin{equation}
\label{qed1}
\Delta E_{QED}(2P)=\frac{\alpha((Z-1)\alpha)^4}{8\pi }\frac{M_e^3}{m_e^2}
\Bigl[-\frac{4}{3}\ln k_0(2P)-\frac{m_e}{6M_e}
\Bigr].
\end{equation}

Relativistic corrections of orders $O(\alpha^4)$, $O(\alpha^6)$
connected with the motion of the electron are also known in analytical form \cite{egs}.
They can be important only for the interval $(2S-1S)$.
As we noted at the beginning of this work, the electron Lamb shift in muon-electron 
ions of lithium, beryllium and boron appears already in the order $ O(\alpha^2) $ 
\eqref{eq23}, but the main term is also proportional to $a_1^2$ .
Thus, \eqref{eq23} contains two small expansion parameters. Although the QED corrections 
\eqref{qed}, \eqref{qed1} have a high order of smallness in $ \alpha$, they 
also depend on the nuclear charge $Z$, which leads to an increase in this correction 
when passing to ions with large $Z$. As a result, with an increase in $Z$, the effect 
of compensation for two corrections is observed. If for the lithium ion the value 
of the shift $ (2P-2S)$ is positive, then already for the boron ion it becomes 
negative, and for the beryllium ion the QED correction and \eqref{eq23} are 
almost completely cancelled. The obtained total values of the energy intervals 
$(2P-2S)$ and $(2S-1S)$ in Tables~\ref{tb1}, \ref{tb2} can serve 
as a reference point for comparison with future experimental data.
They are presented with an accuracy of 3 digits after the decimal point according to the obtained 
analytical formulas. Nevertheless, the calculation accuracy is not so high, since a number of approximations 
are used in the work to calculate the recoil effects $M_e/M_\mu$.
We can estimate the contribution of some of the unaccounted terms to the Lamb shift (2P-2S)
connected with this expansion to 0.5 GHz.

\begin{table}[b]
\caption{Various contributions to the energy intervals $\Delta E(2S\!-\!1S)$ and $\Delta E(2P_{1/2}\!-\!2S)$
in the ($\mu e{\,}^7_3$Li) ion (in GHz) in the variational method.}
\label{varLi}
\begin{center}
\begin{tabular}{|@{\hspace{2mm}}l@{\hspace{3mm}}|@{\hspace{3mm}}r@{\hspace{4mm}}|@{\hspace{2mm}}r@{\hspace{6mm}}|}
\hline\hline
   & $\Delta E(2S\!-\!1S)$ & $\Delta E(2P_{1/2}\!-\!2S)$ \\     \hline
$E_{\rm NR}$ & 9\,869\,022.575  &  36.568 \\   \hline
$\Delta E_{\rm BP}$ &  482.067  & 116.833 \\   \hline
$\Delta E_{\rm fs}(2P_{1/2})$ &  ---       & $-$117.019 \\   \hline
$\Delta E_{\mu N}$  & $-$0.649 & $-$0.056 \\ \hline
$\Delta E_{\rm VP}$ &    3.038  &   0.434 \\    \hline
$\Delta E_{\rm QED}$ & $-$95.367 & $-$14.257 \\    \hline
$\Delta E_{\rm HO\mbox{-}QED}$&  $-$1.56(2)\hspace*{-2.5mm} & $-$0.221(2)\hspace*{-4.5mm} \\    \hline
$E_{tot}$    & 9\,869\,410.10(2)\hspace*{-2.5mm} & 22.160(2)\hspace*{-4.5mm} \\      \hline\hline
\end{tabular}
\end{center}
\end{table}

\section{Explicit three-body formalism}

The obtained results can be improved using the variational method in the three-body problem.
Numerically the nonrelativistic Schr\"odinger solution may be obtained via variational approach with
almost arbitrary precision \cite{Korobov00}. The variational approach allows also to calculate different
corrections by the perturbation theory. Let us focus on the calculation of relativistic corrections
which are equal zero in a two-body approximation.

Relativistic corrections are determined by the following Breit-Pauli Hamiltonian
\cite{BS,Korobov06}:
\begin{equation}
\begin{array}{@{}l}\displaystyle
H_{\rm BP} = -\frac{p_e^4}{8m^3}+\frac{1}{8m^2}4\pi\Bigl(Z\,\delta(\mathbf{x}_{e})\!-\!\delta(\mathbf{x}_{\mu e})\Bigr)
\\[3mm]\displaystyle\hspace{15mm}
+\frac{Z}{2}\>
\frac{p_e^i}{m_e}\left(
\frac{\delta^{ij}}{r_1}+\frac{r_1^ir_1^j}{r_1^3}
\right)\frac{P_N^j}{M}-\frac{1}{2}\>\frac{p_e^i}{m_e}\left(
\frac{\delta^{ij}}{x_{\mu e}}+\frac{x_{\mu e}^ix_{\mu e}^j}{x_{\mu e}^3}
\right)\frac{p_{\mu}^j}{m_{\mu}},
\end{array}
\end{equation}
where $\mathbf{P}_N=-(\mathbf{p}_\mu+\mathbf{p}_e)$ and $\mathbf{x}_{\mu e}=\mathbf{x}_e\!-\!\mathbf{x}_{\mu}$.

\begin{table}[b]
\caption{Energy intervals $\Delta E(2S\!-\!1S)$ and $\Delta E(2P_{1/2}\!-\!2S)$ for muonic He, Li, Be, and B ions
(in GHz) in the variational method.}
\label{varTot}
\begin{center}
\begin{tabular}{|@{\hspace{3mm}}l@{\hspace{3mm}}|@{\hspace{3mm}}r@{\hspace{3mm}}|@{\hspace{3mm}}r@{\hspace{3mm}}|}
\hline\hline
   & $\Delta E(2S\!-\!1S)$ & $\Delta E(2P_{1/2}\!-\!2S)$ \\     \hline
($\mu e{\,}_{2}^{4}$He)    &  2\,467\,150.79(1) &  9.665(1) \\  \hline
($\mu e{\,}_{3}^{7}$Li)    &  9\,869\,410.10(2) & 22.160(2) \\  \hline
($\mu e{\,}_{4}^{9}$Be)    & 22\,207\,596.32(8) &  4.022(8) \\  \hline
($\mu e{}_{\;\;5}^{11}$B)  & 39\,483\,388.3(4)\hspace*{1.8mm} & $-$82.03(4)\hspace*{2.1mm} \\
\hline\hline
\end{tabular}
\end{center}
\end{table}

For the $2P$ state we have as well the fine structure contribution to the Hamiltonian:
\begin{equation}
\Delta H_{\rm fs} = b_{\rm fs}\>(\mathbf{s}_e\!\cdot\!\mathbf{L}),
\end{equation}
where $\mathbf{L}$ is an operator of the total orbital momentum of the three-body system and $b_{\rm fs}$ is expressed in terms of the reduced matrix elements of the three-body operators as follows
\[
\begin{array}{@{}l}\displaystyle
b_{\rm fs} =
   (2\,\mbox{Ry})\frac{\alpha^2}{\sqrt{L(L+1)(2L+1)}}
\\[4mm]\displaystyle\hspace{10mm}
   \times\Biggl\{
   \frac{Z(1\!+\!2a_e)}{2m^2_{e}}\left\langle L\left\|
      \frac{\mathbf{x}_{e} \!\times\! \mathbf{p}_e}{x^3_{e}}
   \right\|L\right\rangle
   +\frac{Z(1\!+\!a_{e})}{Mm_{e}}\left\langle L\left\|
      \frac{\mathbf{x}_{e} \!\times\! \mathbf{P}_{N}}{x^{3}_{e}}
   \right\|L\right\rangle
\\[5mm]\displaystyle\hspace{10mm}
   -\frac{1\!+\!2a_e}{2m^2_e}\left\langle L\left\|
      \frac{\mathbf{x}_{\mu e}\!\times\!\mathbf{p}_e}{x^3_{\mu e}}
   \right\|L\right\rangle\!
   -\!\frac{1\!+\!a_e}{m_{\mu}m_e}\left\langle L\left\|
      \frac{\mathbf{x}_{\mu e}\times\mathbf{p}_{\mu}}{x^{3}_{\mu e}}
   \right\|L\right\rangle
   \Biggr\}.
\end{array}
\]
Results of numerical computation of relativistic and other corrections for the ($\mu e{\,}^7_3$Li) 
ion are summarized in Table \ref{varLi}.

The largest contribution to the leading order QED (one-loop) corrections is the vacuum polarization, Eq.~(\ref{vp2}), due to small subsystem ($\mu N$). For the case of ($\mu e{\,}^7_3$Li) ion $\Delta E_{vp}^{\mu N} = -14\,814\,915$ GHz. Still the total contribution of this leading VP correction to the $2S\!-\!1S$ energy interval is smaller then 1 GHz, since the wave function of the small subsystem differs too little between the states under consideration.

The other QED contributions can be calculated using simplified two-body approximation with a "pseudo-nucleus" ($\mu N$) by means of formulas (\ref{vp1}), (\ref{vp3}) for the vacuum polarization, and Eqs.~(\ref{qed})-(\ref{qed1}) for the one-loop self-energy. Higher-order QED contributions are also important and have to be included. The following one- and two-loop corrections \cite{YS} have been taken into consideration:
\begin{equation}\label{1-loop}
\begin{array}{@{}l}\displaystyle
\Delta E_{\rm 1-loop}(nS) =
   \frac{\alpha(Z^*\alpha)^5}{\pi n^3}
      \biggl[4\pi\left(\frac{139}{128}-\frac{1}{2}\ln{2}+\frac{5}{192}\right)
\\[3mm]\displaystyle\hspace{30mm}
   +(Z^*\alpha)\left(-\ln^2(Z^*\alpha)^{-2}+A_{61}(nS)\ln(Z^*\alpha)^{-2}\right)\biggr]
   +\dots,
\\[3mm]\displaystyle
\Delta E_{\rm 1-loop}(2P_{1/2}) =
   \frac{\alpha(Z^*\alpha)^6}{\pi n^3}\biggl[\frac{103}{180}\biggr]\ln(Z^*\alpha)^{-2},
\\[3mm]\displaystyle
\Delta E_{\rm 2-loop}(nS) =
   \frac{\alpha^2(Z^*\alpha)^4}{\pi^2 n^3}\bigl[0.538941\bigr],
\end{array}
\end{equation}
where $Z^*=Z\!-\!1$ is a charge of a "pseudo-nucleus", the state-depended coefficient $A_{61}(nS)$ is taken as in \cite{YS}, Eq.~(2.5). The recoil effects are small and may be neglected.

Final results of calculations in the three-body formalism are presented in Table~\ref{varTot}. The uncertainties indicated stem from two sources: the uncalculated higher-order contributions and due to imperfectness of the point-like model for the "pseudo-nucleus".

\section{Conclusion}

In this work we investigate the energy levels in muon-electron-nucleus three-particle system.
We calculate the electron energy interval $(2S-1S)$ and electron Lamb shift $(2P-2S)$
in electron-muon ions of lithium, beryllium and boron using the analytical perturbation theory
method and the variational approach. The results of the calculation in analytical perturbation
theory are presented in Tables~\ref{tb1}, \ref{tb2}. The results of the calculation on the basis of
variational method are presented in Tables~\ref{varLi}, \ref{varTot}. The results obtained by different methods
are in agreement with the account of theoretical errors. We investigate the dependence
of the electron Lamb shift $(2P-2S)$ on the nucleus charge Z. An interesting effect of level $2P$ ans $2S$
reorientation is discovered when passing from a beryllium ion to a boron ion.
The extension of studies of the energy levels of electron-muonic helium in \cite{strasser} to three-particle
systems with other nuclei could facilitate the experimental study of this issue.

\acknowledgments
This work is supported by the Russian Science Foundation (grant 18-12-00128).
The work of F.A. Martynenko is supported by the Foundation for the
Advancement of Theoretical Physics and Mathematics "BASIS" (grant No. 19-1-5-67-1).

\end{document}